\documentclass{vldb}
\usepackage[utf8]{inputenc}
\usepackage{algorithm}
\usepackage{algpseudocode,algorithm}
\usepackage{tabularx}
\usepackage{url}
\usepackage{balance}
\usepackage{fouriernc,amsmath} 	
\usepackage{comment}

\newcolumntype{L}[1]{>{\raggedright\let\newline\\\arraybackslash\hspace{0pt}}m{#1}}
\newcolumntype{C}[1]{>{\centering\let\newline\\\arraybackslash\hspace{0pt}}m{#1}}
\newcolumntype{R}[1]{>{\raggedleft\let\newline\\\arraybackslash\hspace{0pt}}m{#1}}

\begin{document}
	
\title{Constellation Queries over Big Data}

\numberofauthors{7} 	

\author{
\alignauthor 	Fabio Porto\\
    \affaddr{LNCC} \\
	\affaddr{DEXL Lab}\\
	\affaddr{Petropolis, Brazil}\\
	\email{fporto@lncc.br}	
\alignauthor Amir Khatibi\\
	\affaddr{UFMG} \\
	\affaddr{ Brazil}\\
	\email{amirkm@dcc.ufmg.br} 
\alignauthor 	Joao Guilherme Nobre\\
	\affaddr{LNCC} \\
	\affaddr{DEXL Lab}\\
	\affaddr{Petropolis, Brazil}\\
	\email{joanonr@lncc.br}
\and    
\alignauthor Eduardo Ogasawara\\
	\affaddr{CEFET-RJ}\\
	\affaddr{Rio de Janeiro, Brazil}\\
	\email{eogasawara@ieee.org}	
\alignauthor Patrick Valduriez\\
	\affaddr{INRIA}\\
	\affaddr{Montpellier, France}\\
	\email{patrick.valduriez@inria.fr}	
\alignauthor Dennis Shasha\\
	\affaddr{New York University}\\
	\affaddr{New York, USA}\\
	\email{shasha@courant.nyu.edu}
}    
\additionalauthors{
     Alberto Krone-Martins, University of Lisbon, Lisbon, Portugal, algol@sim.ul.pt }   

\maketitle

\begin{abstract}
A geometrical pattern is a set of points with all pairwise distances (or, more generally, relative distances) specified.  Finding matches to such patterns has applications to spatial data in  seismic, astronomical, and transportation contexts. For example, a particularly interesting geometric pattern in astronomy is  the Einstein cross, which is an astronomical phenomenon in which a single quasar is observed as four distinct sky objects (due to gravitational lensing) when captured by earth telescopes. Finding such crosses, as well as other geometric patterns, is a challenging problem as the potential number of sets of elements that compose shapes is exponentially large in the size of the dataset and the pattern. In this paper, we denote geometric patterns as constellation queries and propose algorithms to find them in large data applications. Our methods combine quadtrees, matrix multiplication, and unindexed join processing to discover sets of points that match a geometric pattern within some additive factor on the pairwise distances. Our distributed experiments show that the choice of composition algorithm (matrix multiplication or nested loops) depends on the freedom introduced in the query geometry through the distance additive factor. Three clearly identified blocks of threshold values guide the choice of the best composition algorithm. Finally, solving the problem for relative distances requires a novel continuous-to-discrete transformation. To the best of our knowledge this paper is the first to investigate constellation queries at scale.  
\end{abstract}


\section{Introduction}

The availability of large datasets in science, web and mobile applications enables new interpretations of natural phenomena and human behavior. 

\begin{figure}[!ht]
	\centering\includegraphics[width=0.65\linewidth]{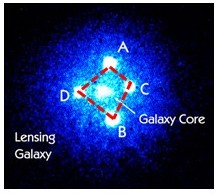}
	\caption{Einstein Cross identification from astronomic catalogs {\protect\cite{nasas_hubble_space_telescope_first_1990}}}
	\label{fig:1}
   	\vspace{8pt}
\end{figure}

 Consider the following two use cases:

\textbf{Scenario 1. }An astronomy catalog is a table holding billions of sky objects from a region of the sky, captured by telescopes. An astronomer may be interested in identifying the effects of \emph{gravitational lensing} in quasars, as predicted by Einstein's General Theory of Relativity \cite{EinsteinRelativity_2015}. According to this 
theory, massive objects like galaxies bend light rays that travel near them just as a glass lens does. Due to this phenomenon, an earth telescope would receive 
two or more virtual images of the
lensed quasar leading to a composed new object (Figure \ref{fig:1}), such as the Einstein cross \cite{Einstein_cross_2015}.

\begin{figure}[!ht]
	\centering\includegraphics[width=0.75\linewidth]{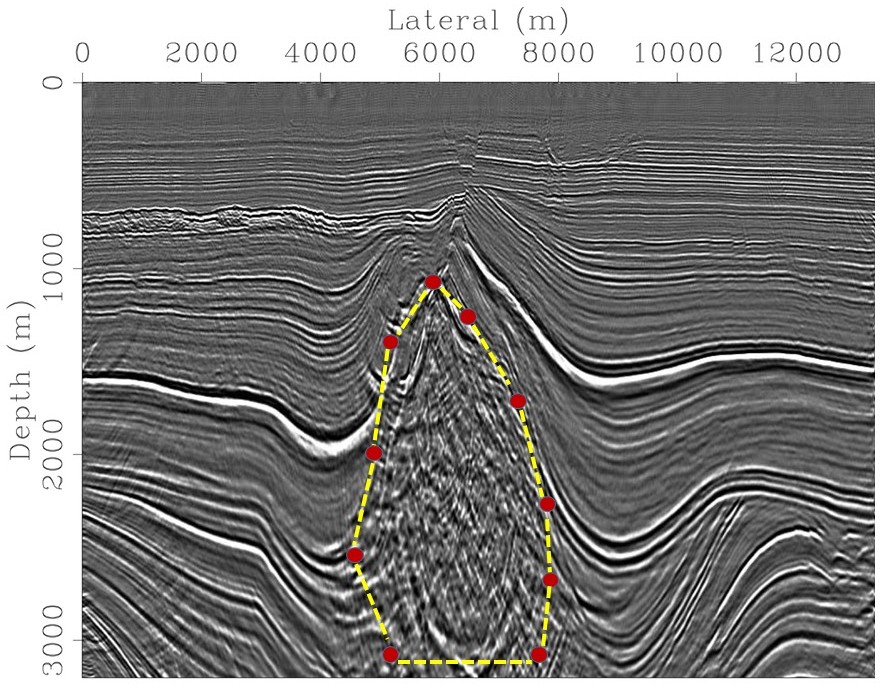}
	\caption{Salt Dome identification from seismic datasets {\protect\cite{sethian_seismic_2006}}}
	\label{fig:2}
   	\vspace{8pt}
\end{figure}

\textbf{Scenario 2.} In seismic studies \cite{three-dimensionalseismicdata_2004}, a huge dataset holds billions of seismic traces, which, for each position  in space, present a list of amplitudes of a seismic wave at various depths (i.e. seismic traces). A seismic interpreter tries to extract meaning out of such datasets by finding higher-level seismic 
objects (Figure \ref{fig:2}) such as: faults \cite{ciarlini_methods_2015}, salt domes, etc. Those \emph{features} may be obtained from the seismic dataset through a spatial composition of seismic traces. Indeed, such compositions convey meaning to the user in terms of real seismic objects of interest. 

In the scenarios above, constellations, such as the Einstein cross or the salt dome, are obtained from compositions of individual elements in large datasets in some spatial arrangement with respect to one another. Thus, extracting constellations from large datasets entails matching geometric pattern queries against sets  of individual data observations, such that each set obeys the geometric constraints expressed by the pattern query.

Solving a constellation query in a big dataset is hard due to the sheer number  of possible compositions from billions of observations. In general, for a big dataset $D$ and a number $k$ of elements in the pattern, an upper bound for candidate combinations $\binom{|D|} {k}$ is the number of ways to choose k items from $D$. This becomes even more challenging with the useful, but more general query which is to find compositions obtained from $D$ that exhibit a similar shape to a query but at a different spatial scale. For example, the user may look for a square at smaller or larger scales.  This paper focuses primarily on {\em pure constellation queries} (when all pairwise distances are specified up to an additive factor), but will show how to extend this to {\em General constellation queries} (where only relative distances are specified). Our running example will come from astronomy.

To become computationally tractable, when $\binom{|D|} {k}$ is big, we propose a strategy to process pure constellation queries that involves three main techniques. First, we use query element properties to constrain candidate sets. Query element properties are element-specific properties that an element must have to be responsive to a query. For stars, this might be a range of light magnitudes in the \emph{Einstein Cross} scenario. Second, we enable the query to be solved at different levels of approximation. At each approximation level, the dataset $D$ is reduced to a set of equivalence classes each consisting of a set of spatially close neighboring elements. Each equivalence class is itself represented by a single point.  At finer levels of granularity, the equivalence classes shrink.  Third, to handle errors in measurement,  we allow matches up to an additive factor $\epsilon$. For example, two stars might be 100 million kilometers distant +/- 10 million kilometers. Fourth, after pairwise distances have been evaluated, the distances must be composed to see whether the global constraints of the query are satisfied. For example, to look for a triangle having side lengths L1, L2, and L3, we want to find points a, b, c such that the distance between a and b is of distance L1 (within an $\epsilon$ additive factor), and similarly for b and c and for c and a. We make use of two classes of composition algorithms for this step: Bucket\_NL and Matrix Multiplication. 




The remainder of this paper is organized as follows. Section \ref{sec:usecase} presents a use case based on the Einstein cross, from an astronomy dataset.  Section \ref{sec:problem_formulation} formalizes the constellation query problem. Section \ref{sec:constellation_query_processing} presents our techniques to process constellation queries. In section \ref{sec:compositionalgorithm}, we preset our algorithms. Next, section \ref{sec:general_constellation_queries} presents a theory for general constellation queries. Section \ref{sec:experimental_evaluation} discusses our experimental environment and discusses the evaluation results, followed by section \ref{sec:related_works} that discusses related work. Finally, section \ref{sec:conclusion} concludes.

\section{Use case: The Einstein Cross in Astronomy}
\label{sec:usecase}

Astronomical surveys collect data from regions of the sky by means of some capturing instrument, such as an optical telescope. From the obtained digital images, sky objects are identified in a reduction pipeline, and registered in a large table of sky objects, named the sky catalog.


Data from the well-known Sloan Digital Sky Survey (SDSS)\footnote{http://skyserver.sdss.org/dr12/en/help/browser/browser.aspx} catalog can be modeled as the following relation:

\begin{equation*}
	{SDSS (Obj\_ID, RA, DEC, u, g, r, i, z, Redshift, \ldots} ) 
	\label{eq:1}
   	\vspace{8pt}
\end{equation*}

The attributes $u$, $g$, $r$, $i$, $z$ refer to the magnitude of light emitted by an object measured at specific wavelengths. Their values are measured in logarithmic units, through various wavebands, from ultraviolet to infrared. A constellation in the SDSS scenario would be defined by a sequence of objects from the catalog whose spatial distribution forms a shape. The Einstein Cross, also known as \emph{Q2237+030} or \emph{QSO2237+0305},  presents a cruciform spatial configuration. This Einstein cross is observed as four images of a single distant quasar whose emitted light bends when passing by a much closer galaxy, producing a visual effect of the cross when the image reaches the earth, see points: A, B, C, and D in Figure \ref{fig:1}. The observation of the Einstein cross led to the confirmation of Einstein's Theory of General Relativity and can also indicate the presence of dark matter.

In a constellation query, an astronomer may use a specific Einstein Cross as a query. The latter would include a sequence of objects (A, B, C, and D) as its elements with attributes ra, dec, magnitude, etc. In this context, the query would request to find sequences of sky objects with similar spectra and conveying a shape close to that exhibited by the sample. Indeed, other  shapes have been reported \cite{1538-4357-646-1-L45} in the literature and these could be found by running a shape constellation query.

\section{Problem Formulation}
\label{sec:problem_formulation}

In this section, we introduce the problem of answering pure Constellation Queries on a dataset of  objects. A Dataset $D$ defined as a set of elements (or objects) $D= \{e_1,e_2,\ldots,e_n\}$, in which each e$_i$, $1 \le i \le n$, is an element of a domain $Dom$. Furthermore, $e_i = <atr_1, atr_2, \ldots ,atr_m>$, such that $atr_j$ ($1 \le j \le m$) is a value describing a characteristic of $e_i$. 

A constellation query $Q_k= \{q_1,q_2,\ldots,q_k\}$ is (i) a sequence of $k$ elements of domain $Dom$, (ii) the distances between the centroids of each pair of query elements that define the query shape and size with an additive allowable factor $\epsilon$, and (iii) an element-wise function $f(e,q)$ that computes the similarity (e.g. in brightness at a certain wavelength) between elements $e$ and $q$ up to a threshold $\theta$. 

For example, for an Einstein cross at a particular scale, the query size  is four and the distance constraints among points A, D, B, C are approximately as follows: $d(A,D)~=~2.748 \cdot 10^{-5},\ d(D,B)~=~2.624 \cdot 10^{-5},\ d(B,C)~=~2.148 \cdot 10^{-5}, \\ d(A,C)~=~9.201 \cdot 10^{-6}, d(C,D)~=~3.437 \cdot 10^{-5},\ d(A,B)~=~2.273 \cdot 10^{-5}$. The element function $fe$ is an attribute value comparison for each Einstein cross element.

A sequence $s$ of elements of length $k$  in $D$ {\em property matches} query $Q$ if every element $s[i]$ in $s$ satisfies $fe(s[i], q_{i})$ up to a threshold $\theta$ and for every $i,j \le k$: (i) the distance between elements $s[i]$ and $s[j]$ is within an additive factor $\epsilon$ of the distance between $q_i$ and $q_j$, which is referred to as \emph{distance match}. The solutions obtained using \emph{property match} and \emph{distance match} to solve a query $Q$ are referred to as \emph {pure constellations}. 
In a \emph{general constellation query}, a sequence $s$ of elements of length $k$  in $D$ matches  query $Q$ if every element $s[i]$ in $s$ satisfies $fe(s[i], q_{i})$ up to a threshold $\theta$ and there exists a $c$ such that for every $i,j \le k$: (i) $c$ times the distance between elements $s[i]$ and $s[j]$ is within an additive factor $ \epsilon$ of the distance between $q_i$ and $q_j$.


\section{Pure Constellation Queries}
\label{sec:constellation_query_processing}

Applying \emph{pure constellation} to find patterns such as the Einstein cross over an astronomy catalog requires efficient query processing techniques as the catalog  may hold billions of sky objects. The 2MASS catalog \cite{Skrutskie:06}, for instance, holds 470 million objects, which would lead to the evaluation of roughly $\frac{470M^{4}}{{4}!} =  2.0 \cdot 10^{33}$ candidate sets. 

In this context, efficiently answering pure constellation queries involves constraining the huge space of candidate sets.

We adopt the following processing strategies:

\begin{itemize}
 
\item Index the data to prune clusters of stars that do not contribute to solutions taking the error bound $\epsilon$ into account.
\item After filtering, keep candidate sets that respect the query distances.
\end{itemize}




The next sections describe in detail the query processing techniques.

\subsection{Reducing Data Complexity \  using a Quadtree} 
\label{sec:queryapprox}
A constellation query looks for patterns in large datasets, such as the 2MASS catalog. Computing constellation queries involves matching each star to all neighboring stars  with respect to the distances in the query, a costly procedure in large catalogs. To reduce this cost, we adopt a filtering process that eliminates space regions where solutions cannot exist.  

\begin{figure}[!ht]
    	\centering\includegraphics[width=1.0\linewidth]{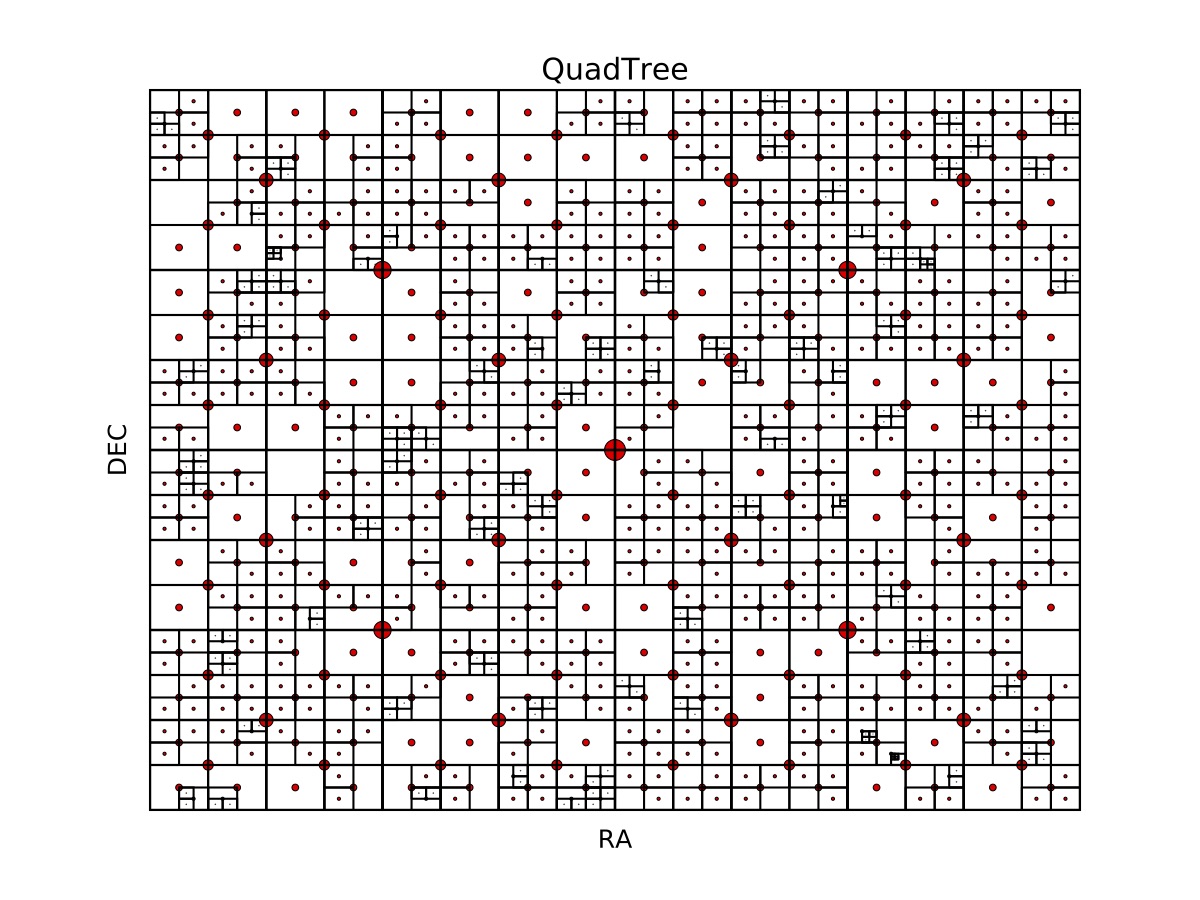}
	\caption{Quadtree structure for a dense region }
	\label{fig:quadtreedatadistribution}
   	\vspace{8pt}
\end{figure}

The filtering process is implemented on top of a quadtree  \cite{Samet_1990}, constructed over the entire input dataset. The quadtree splits the 2-dimensional catalog space into successively refined quadrangular regions. Figure \ref{fig:quadtreedatadistribution} depicts a representation of a quadtree holding data from the 2Mass catalog. The horizontal and vertical axes correspond to the polar coordinate (RA and DEC), respectively.  

\begin{figure}[!ht]
	\centering\includegraphics[width=1.0\linewidth]{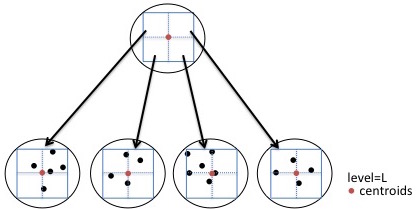}
	\caption{Quadtree node representation}
	\label{fig:quadtree}
   	\vspace{8pt}
\end{figure}

A node in the tree is associated to a spatial quadrant. The geometric center of the quadrant is the node centroid and is used as a representative for all stars located in that quadrant, for initial distance matching evaluation. The quadtree data structure includes: a root node, at level $L=0$; a list of intermediary nodes, with level $ 1 \le L \le tree\_height -1$; and leaf nodes, at level $ L= tree\_height$. To avoid excessive memory usage, data about individual stars are stored only in  leaf nodes, Figure \ref{fig:quadtree}.

Query answering using the quadtree reduces the data complexity by restricting star matching operations to spatial regions where distance constraints have some chance of being satisfied. 

The algorithm begins by determining the level of the quadtree $L_{e}$ at which the $\epsilon$ error bound exceeds the diameter of the node. If we make the reasonable assumption that $\epsilon$ is less than the minimum distance between elements in the query $(minq)$, then at height $L_{e}$ no two stars would be covered by a single quadtree node. 

\begin{figure}[!ht]
	\centering\includegraphics[width=\linewidth]{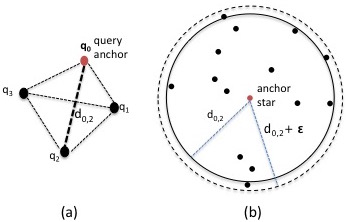}
	\caption{(a) Pure constellation query with anchor and maximum distance (b) Neighboring elements of anchor element}
	\label{fig:geometryquery}
   	\vspace{8pt}
\end{figure}

Given a star s that will correspond to the centroid $q_0$ of the pattern being matched, the first step is to eliminate all parts of the quadtree that could not be relevant. The algorithm finds  the node at level $L_{e}$ containing s. That is called the query anchor node. The algorithm finds the nodes that lie within a  radius $\rho$ of the query anchor node, where $\rho$ is  the maximum distance plus the additive error bound $\epsilon$ between the centroid of the query pattern and any other query element. As depicted in Figure \ref{fig:geometryquery}, in (a) a query Q has an anchor element $q_0$ and the largest distance to the remaining query elements $d_{0,2}$. In (b), a star is picked as an anchor and all neighboring stars within distance $d_{0,2}+ \epsilon$ are preliminary candidates for distance matching. 

Next the algorithm determines for each pattern element $q_i$, which stars are at distance dist($q_0$, $q_i$) from s within an additive factor of $\epsilon$ (see algorithm \ref{alg:quadtree} Function FindMatchingStars). Such stars might correspond to $q_i$ in the case where s corresponds to $q_0$. For each pair of nodes n1 and n2, where n1 contains s and n2 may contain stars that correspond to $q_i$ for some $i$,  the algorithm checks whether the  distance between the centroids of n1 and n2 matches dist($q_0$, $q_i$), taking into account both the diameter of the nodes and the error bound. This procedure filters out all node pairs at that level for which no matching could possibly occur. 

If n2 has not been filtered out, then a simple test determines whether going one level down the tree is cost effective as opposed to testing pairs of individual stars in the two nodes. That test consists of determining whether any pair of children of n1 and n2 will be eliminated from consideration based on a distance test to the centroids of those children. If so, then the algorithm goes down one level (see algorithm \ref{alg:quadtree}, \emph{Function TreeDescend}). The same matching procedure is applied to the children nodes of n1 and n2 respectively. If not, then the stars for n1 and n2 are fetched and any star in n2 that lies within an $\epsilon$ additive bound of s is put into bucket $B_i$.

\begin{algorithm}[!ht]
	\caption{Pure constellation filtering} 
	\label{alg:quadtree}
    
\vspace{8pt}
\vspace{8pt}

$\textbf{Function\ FilterNeighbors}(Query\ q, Quadtree\\ qt,Element\ e, list \  listNode,fe\_predicate\ fe, fe\_threshold\ \theta)$
		
		\begin{algorithmic}[1]
			\State $q_{0} \gets q.anchor()$
			\State $ne \gets \emptyset$
			\State $max \gets q.maxDistance()$
			\If {$fe (q_{0},e) \le \theta $}
				\State {$ne \gets\ qt.neighbors(e,max, \epsilon)$}
				\State {$ne \gets ne \cap listNode$}
			\EndIf
			\State $return(ne)$
			\newline
			\end{algorithmic}

\vspace{8pt}

               $\textbf{Function\ FindMatchingStars}(Bucket []\ b, Query\ q, \\ ~~~~ Element\ e, Element\ s, fe\_predicate\ fe, fe\_threshold\ \theta, \\ ~~~~ dist\_threshold\ \epsilon, Matrix \\ distMatrix)$
		
		\begin{algorithmic}[1]
			\State $q_{0} \gets q.anchor()$
			\For {$ q_{i}\ in\ q_0.neighbors() $}
				\State $ dist \gets  EuclidianDistance(e,s) $
				\State $ distQ \gets distQuery(0,i) $	
				\If {$(distQ - \epsilon \le dist \le distQ + \epsilon)$}                        
                   \State {$ matchList \gets TreeDescend(e,\ s,\ distQ,\  q_i,\  \epsilon)$}
                   \For {$ ms \in matchList$}
						\State{$b[q_i] \gets [ms[0],ms[1]]$}
                    \EndFor   
				\EndIf
			\EndFor
			\State $return(b)$	
			\newline
		\end{algorithmic}

\vspace{8pt}

         $\textbf{Function\ TreeDescend}(Element\ n_1, Element\ n_2, Distance \ dist,\\ QueryElement\  q_i, dist\_threshold\  \epsilon)$
		
		\begin{algorithmic}[1]
			\State $nonpairs \gets findNonPairs(n_1,n_2,dist,\epsilon)$
			\If {$(nonpairs \neq \emptyset) $}
            	 \State {$listNode_1 \gets n_1.splitNode() $}
                \State {$listNode_2 \gets n_2.splitNode() $}
				\State {$candidates \gets\ produceCandidates(listNode_1,\ $
                   $listNode_2, \ nonpairs)$}
				\For {$cpair \in candidates$}
					\State {$ TreeDescend (cpair[0],cpair[1], dist) $}
				\EndFor
            \Else
            	\State {$ starPairsList \gets produceStarPairs(n_1,\ n_2,\  dist,\ q_i,\  \epsilon) $}
			\EndIf
			\State $return(starPairList)$
			\newline
			\end{algorithmic}

\vspace{8pt}
\vspace{8pt}

\end{algorithm}

\subsection{Composition algorithms}
In this section, we discuss approaches to join the buckets produced by the  filtering step. As we will observe in section \ref{sec:experimental_evaluation}, composition algorithms are the most time consuming operation in processing constellation queries. A given anchor node may generate buckets containing thousands of elements. Thus, finding efficient composition algorithms is critical to efficient  overall processing.

\subsubsection{Buckets Nested Loop (Bucket-NL)}
\label{sec:bnl}
An intuitive way to produce constellations for a given anchor element is by directly joining the buckets of candidate elements considering the corresponding pairwise distances between query elements as the join predicate. In this approach, each bucket is viewed as a relation, having as a schema their spatial coordinates and an id, $B_{i}(starid,ra,dec)$. A solution is obtained whenever a tuple is produced having one neighbor element from each bucket, such that the distances between each element in the solution \emph{distance-match} those among respective query elements, +/- $\epsilon$. Bucket-NL assumes a nested loop algorithm to traverse the buckets of candidate elements and checks for the distance predicates. Thus, applying a \emph{distance-match} constraint corresponds to applying a cyclic join among all buckets in the bucket set followed by a filter among non-neighbors in the cycle.  For example, Bucket-NL would find pairs (t1, t2) where t1 is from $B_i$ with and t2 from $B_{i+1}$ if dist(t1, t2) is within dist($p_i$, $p_{i+1}$) +/- $\epsilon$. Then given these pairs for buckets 1 and 2, buckets 2 and 3, buckets 3 and 4, etc, Bucket-NL will join these cyclically and then for any k-tuple of stars s1, s2, ..., sk that survive the join, Bucket-NL will also check the distances of non-neighboring stars (e.g. check that dist(s2,s5) = dist($p_2$, $p_5$) +/- $\epsilon$).

\subsubsection{Matrix Multiplication based approaches}
 
The Matrix Multiplication $(MM)$ based approaches precede the basic $Bucket\_NL$ algorithm by  filtering out candidate elements. Here are the details: recall that bucket $B_{i}$ holds elements for the candidate anchor that correspond to dist($q_0$, $q_i$) +/- $\epsilon$. Compute  the  matrices: $M1(B_{1}, B_{2}), M2(B_{2}, B_{3}), M3(B_{3}, B_{1})$ where $Mi(B_{i}, B_{i+1})$ has a 1 in location j, k if the jth star in $B_i$ and the kth star in $B_{i+1}$ is within dist($p_i$, $p_{i+1}$) +/- $\epsilon$. The product of matrices indicates the possible existence of solutions for a given anchor element, as long as the resulting matrix contains at least a one in its diagonal (see algorithm \ref{alg:MMF}). 
The MM approach can be implemented with fast matrix multiplication algorithms \cite{FastMM1}\cite{FastMM2} and enables quick elimination of unproductive bucket elements.

\subsubsection{MMM Filtering}
Matrix multiplication may be applied multiple times to eliminate stars that cannot be part of any join. The idea is to apply  $k$ matrix multiplications, each with a sequence of matrices starting with a different matrix (i.e.  a $B_{i}$ bucket appears in the first and last matrices of a sequence, for $1 \leq i \leq k$). The resulting matrix diagonal cells having zeros indicate that the corresponding element is not part of any solution and can be eliminated.
For example, for buckets $B_{1}, B_{2}, B_{3}$ and matrices $M1(B_{1}, B_{2}), M2(B_{2}, B_{3}), M3(B_{3}, B_{1})$, we would run $<M1 \cdot M2 \cdot M3>$; $<M2 \cdot M3 \cdot M1>$ and $<M3 \cdot M1 \cdot M2>$. For the multiplication starting with say $M1$, elements in bucket $B_{1}$ with zeros in the resulting matrix diagonal are deleted from $B_1$, reducing the size of the full join.  
 
\subsubsection{Matrix Multiplication Compositions}
The matrix multiplication filtering is coupled with a composition algorithm leading to $MM\_Composition$ algorithms. The choices explore the tradeoff between filtering more
by applying the \emph{MMM} filtering strategy or not.

The \emph{MMM\_NL} strategy uses the \emph{MMM} filtering strategy to identify the elements of each bucket that do not contribute to any solutions and can be eliminated from their respective buckets. Next, the strategy applies \emph{Bucket\_NL} to join the buckets with elements that do contribute to solutions.

The \emph{MM\_NL} considers a single bucket ordering with the anchor node bucket at the head of the list. Thus, once the multiplication has been applied, elements in the anchor node bucket that appear with zero in the resulting matrix diagonal are filtered out from its bucket. Next, the strategy applies  \emph{Bucket\_NL} to join the buckets with anchor elements that do contribute to solutions.

\begin{algorithm}[!ht]
	\caption{MM Filtering} 
	\label{alg:MMF}
    $\textbf{Function\ MM Filtering} \\
    (Query\ q, dist\_threshold\ \epsilon Bucket[] b, Element anchor)$
\begin{algorithmic}    
	\State $M \gets ProducePartialJoin(q, \epsilon, b, anchor)$
	\For {$ m\ in\ M $}
	    \State {$s \gets\ sequence(M,m)$}
	    \State $M_f \gets matrixMultiplication(s)$
	     \If {$ M_f.diagonal() \ne zeros $}
			 \For {$v \  in M_f.diagonal() $}
				 \If {$ v.value == zero$}
					\State {$nonProductive \  \gets \ v.getElement() $}  
		            \State {$b[m].delete(nonProductive)$}
			     \EndIf
			  \EndFor
		\Else 
			\State {$return()$}	     	
	     \EndIf
	\EndFor
	
	\State $Result.add ( Nested\_loop(b,q,dist\_threshold \  \epsilon, anchor))$
	\State $return(Result)$     
	\newline
   \end{algorithmic}
\end{algorithm}
\vspace{8pt}

\subsubsection{Existential Queries}
A particular application of matrix multiplication is the evaluation of existential constellation queries. Such queries test for the existence of a match to the Constellation Query without returning the actual compositions. Such queries can be answered by matrix multiplication alone.

\section{Algorithms for Pure Constellation Queries}
\label{sec:compositionalgorithm}
 
To compute \emph {Pure Constellation Queries}, the overall algorithm implements \emph{property matching} and finds matching pairs, whereas the composition algorithms implement \emph{distance matching} as discussed above.

\subsection{Main  Algorithm}

Algorithm \ref{alg:QP} depicts the essential steps needed to process a Constellation query. The main function is called \emph{ExecuteQuery}. It receives as input a query $q$, dataset $D$, element predicate $fe$, similarity threshold $\theta$, and error bound $\epsilon$. At step 1, a quadtree entry level $L_{e}$ is computed. Next, a quadtree $qt$ is built  covering all elements in $D$ and having height $L_{e}$. Figure \ref{fig:quadtreedatadistribution} illustrates a typical quadtree built on top of  heterogeneously distributed spatial data. The quadtree nodes at level $L_{e}$ become the representatives of stars for initial distance matching. Considering the list of nodes at level $L_{e}$, in line 4, an iteration picks each node, takes it as an anchor node, and searches $qt$ to find neighbors. The geometric centroid of the node quadrant is used as a reference to the node position and neighborhood computation.  Next, each pair (anchor node, neighbor) is evaluated for distance matching against one of the  query pairs: (query anchor, query element) and additive factor $\epsilon$. Matching nodes contribute with stars for distance matching or can be further split to eliminate non-matching children nodes (Function TreeDescend). Matching stars are placed in a bucket holding matches for the corresponding query element, line 8 of function \emph{FindMatchingStars}.

The \emph{Compose Function}, applies a composition algorithm, described in the previous section, between buckets $B=\{B_1,B_2,\ldots,B_k\}$, for $q.size=k+1$, to see which k+1-tuples match the pure constellation query.  The composition algorithm builds a query execution plan to join buckets in $B$,  lines 2 and 9. The distance matching of elements in buckets $B_i$ and $B_j$, $i \ne j$, and $i, j \ne anchor$, is applied by checking their pairwise distances +/- $\epsilon$, with respect to the corresponding distances between $q_i$ and $q_j$, in $q$. 

The choice between running \emph{Bucket\_NL} or a \emph{MM\_filtering} algorithm to implement element composition, as our experiments in section \ref{sec:experimental_evaluation} will show, is related to the size of the partial join buckets. For dense datasets and queries with an error bound  close to the average distance among stars, lots of candidate pairs are produced and \emph{MM\_filtering} improves composition performance, see Figure \ref{fig:high}.   

\vspace{8pt}
	\begin{algorithm}[!ht]
		\caption{Constellation Query Processing} 
		\label{alg:QP}
		
		$\textbf{Function\ ExecuteQuery}(Query\ q, Dataset\ d, 
		\\ ~~~~ fe\_predicate\ fe, fe\_threshold\ \theta, dist\_threshold\\ \epsilon) 
		\\ ~~~~ Bucket[]\ b
		\\ ~~~~ Node[]\  listNodes$
  		
		\begin{algorithmic}[1]
			
            \State $ l_{e} \gets computeTreeHeight(\ qt,\  \epsilon)$
            \State $qt \gets build\_Quadtree(D,l_{e})$
			
			\State $ distMatrix \gets computePairWiseDist(q)$
			
			\State $listNodes \gets qt.getNodesAtLevel(l) $
			
			\For {$ n \ in\ listNodes $}
			     \State $neighbor \gets  FilterNeighbors(q,qt,n,listNode,fe,\theta)$
			     \For {$ne \ in \  neighbor $}
				  	\State {$ b \gets FindMatchingStars(b, q, n, ne, fe, \theta, \epsilon, distMatrix)$}
				 \EndFor 	
			\EndFor
			\State $Result.add ( Compose(b,q,\epsilon))$
			\State $return(Result)$
		\end{algorithmic}
		\end{algorithm}
\vspace{8pt}

\vspace{8pt}
\begin{algorithm}
\caption{Compose Algorithm}
\label{alg:composition}
		$\textbf{Function\ Compose}(Bucket[] b, Query[]\ q, dist\_threshold\ \epsilon)$
		
		\begin{algorithmic}[1]
			\State	{$compositionAlgorithm\ \gets buildPlan(b,q,\epsilon)$} 
			\State {$candidates \gets compositionAlgorithm.next()$}
			\While {$(candidates \neq \ "end")$}
				\For {$ c \ in \ candidates$}
					\If {$ (checkScale(q,c))$}  
						\State	{$\textbf{Solutions}.Add(c)$}
					\EndIf
				\EndFor
				\State {$candidates \gets compositionAlgorithm.next()$}	
			\EndWhile
			\State $return (Solutions)$
		\end{algorithmic}				
	\end{algorithm}
\vspace{8pt}

\section{General Constellation Queries}
\label{sec:general_constellation_queries}

Pure constellation queries specify both the properties and the distances of a pattern and 
require any matching sequence of stars to match the distances +/- $\epsilon$. General 
constellation queries allow s1, ... , sk to match a pattern p1, ... pk if there is some
scale factor $f$ such that dist(si, sj) is within (f $\times$ dist(pi, pj)) +/- $\epsilon$. Because the scale $f$ can take on any real value, there are an uncountable infinity of possible scale factors. 
The challenge is to find a sufficient discrete
set to explore and to do so efficiently. 

\subsection{General Constellation Query Algorithm}

Our basic strategy consists of seven steps:

i)  find the most distant pair of pattern points in the query, which we will denote as p1 and p2, and set their distance, without loss of generality, to be 1.

ii) for every pair of stars s1 and s2 (perhaps after constraining the pair to be not closer
than some threshold and not more distant than some other threshold), 
posit a scale factor {\em scalebasic} to be dist(s1,s2)/dist(p1,p2) and set the error bound $\epsilon$ either to be a constant or to be a function of scalebasic, as dictated by
application considerations.

iii) for a candidate star s to correspond to pattern point pi (for $i$ $\neq$ from 1 or 2),
dist(s1, s) should be within $(scalebasic\  \times dist(p1, pi) )$$ +/- 2 \epsilon$ and 
dist(s2, s) should be within $(scalebasic \times dist(p2, pi) )$$ +/- 2 \epsilon$. Such a star s would then belong to bucket $B_i$. (Note the use of $2 \epsilon$. We discuss this further below.)

iv) Bucket $B_1$ consists of just s1 and bucket $B_2$ consists of just s2.

v) Perform matrix multiplication optionally with respect to all the buckets, just as in the pure constellation query algorithm.

vi) Perform a nested loop join as in the pure constellation query algorithm.

vii) Post-processing: Any sequence of matching stars has to be validated with respect to  an error bound of $\epsilon$ as explained below.

\subsection{Theory and Explanation}

Given an additive error bound of $\epsilon$, our initial search uses an error bound
of $ 2 \epsilon$. 
Here is an example that shows why this might be useful. Suppose the search is for an equilateral triangle consisting of pattern points p1, p2, p3. 
Because this is an equilateral triangle, all intra-pattern distances are the same. So, we can set the maximum intra-pattern element distance to 1 without loss of generality. Suppose further that  $\epsilon = 2$.
If stars s1, s2, s3 have the following pairwise distances dist(s1, s2) = 8, dist(s1, s3) = 12,
dist(s2,s3) = 12, then we might match p1 and p2 to s1 and s2 as candidates.
This would yield scalebasic = 8. Because the error bound through step vi is $2 \epsilon = 4$ and because dist(s2,s3) = (scalebasic $\times$ dist(p2,p3)) + 4, s3 would also be considered a match up to step vi. The post-processing (to be described below) step shows in fact that s1, s2, and s3 are a good match with an error bound
of $\epsilon$  using a scale factor of 10. By allowing the error bound to be $2 \epsilon$, we eliminate the need to test the infinite number of  scale factors between 8 and 12 while still capturing valid matches. Because we may also capture invalid matches using the $2 \epsilon$ bound (we will see examples of this later), 
we need the post-processing step vii above.

Lemma 1 (no false negatives): Suppose  that p1 and p2 are the most distant of the pattern points and dist(p1,p2) = 1. Suppose further there is some matching sequence of stars s1, s2, s3, ..., sk corresponding to a pattern p1, p2, ... pk based on some scale factor $f$ and error tolerance $\epsilon$.  Then s1, s2, s3, ..., sk will be found to correspond to p1, p2, ... pk using a scale factor of scalebasic = dist(s1,s2)/dist(p1,p2) and an error tolerance of $2 \epsilon$. 

Proof: (a) The scale factor $f$ must  allow s1 and s2 to match p1 and p2.  Therefore $(f \times dist(p1,p2)) - \epsilon $ $ \leq (scalebasic \times dist(p1,p2)) $ $ \leq (f \times dist(p1,p2)) + \epsilon $. Because  dist(p1,p2) = 1, this implies that  $|f - scalebasic| \leq \epsilon$, so $scalebasic - \epsilon $ $ \leq f$ $ \leq scalebasic + \epsilon$.  \\
(b) Consider any  distance dist(si, sj). Because si and sj match pi and pj by assumption, $(f \times dist(pi, pj)) - \epsilon $ $\leq dist(si, sj) $ $ \leq $ $ (f \times dist(pi, pj)) + \epsilon$. By (a), $((scalebasic - \epsilon) $ $ \times dist(pi,pj)) - \epsilon $ $ \leq (f \times dist(pi, pj)) $ $ - \epsilon $. Since dist(p1,p2) is maximal, $dist(pi,pj) \leq dist(p1,p2) = 1$, so $(scalebasic \times dist(pi,pj)) $ $ - (2 \epsilon)$ $ \leq \\ ((scalebasic - \epsilon) \times dist(pi,pj)) $ $ - \epsilon$. A similar argument shows that $ (f \times dist(pi, pj)) $ $ + \epsilon $ $ \leq (scalebasic \times dist(pi,pj)) $ $ + (2 \epsilon)$. So, s1, s2, ... sk will be found. QED

Sometimes it would be more natural for an application for the error tolerance $\epsilon$ to increase monotonically with the scale factor. For example,  the error bound $\epsilon$ might increase linearly with the scale factor (e.g. $e \times scale$ for $e < 1$). In such a case, steps (i) through (vi) would set  the $\epsilon$  to be $e \times f$ where $f$ is a scale factor such that $f > scalebasic$ and such that $f - scalebasic = e \times f$. So $(1-e) \times f = scalebasic$ or $f = scalebasic/(1-e) $ and therefore $\epsilon = e \times scalebasic/(1-e) $. For example, if $\epsilon = 0.1 \times scale$ and scalebasic = 10, then for purposes of steps (i) through (vi) set $\epsilon $ to be $0.1 \times 10/0.9 = 1.11111...$, so twice $\epsilon$ is 2.2222.... The goal again is to avoid false negatives, even when we don't know what $f$ could be exactly.

Corollary:  Suppose  that p1 and p2 are the most distant of the pattern points and dist(p1,p2) = 1. Suppose further there is some matching sequence of stars s1, s2, s3, ..., sk corresponding to a pattern p1, p2, ... pk based on some scale factor $f$ and error tolerance $\epsilon$, where $\epsilon = e \times f$, for some $e < 1$.  Then s1, s2, s3, ..., sk will be found to correspond to p1, p2, ... pk using a scale factor of scalebasic = dist(s1,s2)/dist(p1,p2) and an error tolerance of $2 \times scalebasic/(1-e)$.

After searching using $2 \epsilon$, step vii (post-processing) tests whether a given sequence of stars S = s1, s2, s3, ..., sk that corresponds to pattern  P = p1, p2, ... pk using a $2 \epsilon$ error tolerance will correspond to P using an $ \epsilon$ tolerance for some scale factor. To find that scale factor,  for each i, j such that $ 1 \leq i, j \leq k$, determine the minimum and maximum scale factor $minscale_{i,j}$ and $maxscale_{i,j}$ such that
$ dist(si, sj)  =$$ (minscale_{i,j}  $$ \times dist(pi, pj)) + \epsilon  $ and $ dist(si, sj) = $ $ (maxscale_{i,j} \times dist(pi, pj)) $ $ - \epsilon  $. Let the maximum of the minimum minscales be denoted MaxMin and
the minimum of the maxscales be denoted MinMax. If MaxMin $\leq$ MinMax, then any value in the range between MaxMin and MinMax will be a satisfying scale factor. Otherwise there is no satisfying scale factor. Note that this holds whether $\epsilon$ depends on the scale factor or not.

Lemma 2: Post-processing step vii as described in the paragraph above correctly determines whether S corresponds to P based on an $ \epsilon$ tolerance.

Proof: By construction. The minscale and maxscale values are the minimum and maximum possible scale factors for each pair i, j. Any single scale factor $f$ that lies between the minimum and maximum possible scale factors for  all i,j will work. If there is no such scale factor, then none can work. QED

Recall our first example from above:  we are looking for an equilateral triangle (so dist(p1,p2) = dist(p2,p3) = dist(p3, p1) = 1)  and $\epsilon = 2$.
If s1, s2, s3 have the following pairwise distances dist(s1, s2) = 8, dist(s1, s3) = 12,
dist(s2,s3) = 12, then $minscale_{1,2} = 6$, $maxscale_{1,2} = 10$, $minscale_{1,3} = 10$,\\
$maxscale_{1,3} = 14$, 
$minscale_{2,3} = 10$, and
$maxscale_{2,3} = 14$. So MaxMin = 10 and MinMax = 10, so 10 would work.

By contrast, if we are still looking for the same equilateral triangle and $\epsilon = 2$ and we have 
dist(s1,s2) = 6, dist(s2,s3) = 10, dist(s3, s1) = 14, then in the $2 \epsilon$ search,
s1, s2, s3 would correspond to p1, p2, p3 when using a scale factor of 10. But when reduced to $\epsilon = 2$, $minscale_{1,2} = 4$, $maxscale_{1,2}~=~8$, $minscale_{2,3}~=~8$, $maxscale_{2,3} = 12$, $minscale_{1,3} = 12$, and $maxscale_{1,3} = 16$. So MaxMin = 12 and MinMax = 8, thus $MaxMin \leq MinMax$ fails to hold. Therefore, the post-processing step would determine that s1, s2, s3 does not match the pattern.

For an example in which $\epsilon$ is not constant: suppose e = 0.2 and we are looking for an equilateral triangle and dist(s1, s2) = 8, dist(s1, s3) = 12,
dist(s2,s3) = 12.
Then $minscale_{1,2} = 8/1.2 = 6.7$, $maxscale_{1,2} = 10$, $minscale_{1,3} = 10$,
$maxscale_{1,3} = 15$, $minscale_{2,3} = 10$, and
$maxscale_{2,3} = 15$. (To see how these calculations work, consider the computation of $minscale_{1,2}$. We know that $minscale_{1,2} \times (1 + e) = 8$, so $minscale_{1,2} = 6 $ $ 2/3$. Similarly, $maxscale_{2,3} \times (1-e) = 12$, so $maxscale_{2,3} = 12/0.8 = 15$.)
Thus, MaxMin = 10 and MinMax = 10 and s1, s2, s3 would match.

General Constellation Query Theorem: The general constellation algorithm finds all matches of any arbitrary pattern for a given error tolerance $\epsilon$. 

Proof: Lemma 1 tells us that for all scale factors $f$ any sequence of stars that matches the pattern based on $f$ will be found by steps (i) through (vi). Lemma 2 tells us that any sequence of stars found by (i) through (vi) and verified by step (vii) will be correct for the specified error tolerance $\epsilon$. QED.

\section{Experimental Evaluation}
\label{sec:experimental_evaluation}
In this section, we start by presenting our experimental setup. Next, we assess the different components of our implementation for Constellation Queries.
\subsection{Set Up}
\subsubsection{Dataset Configuration}
\label{subsec:dataset_configuration}
The experiments focus on the Einstein cross constellation query and are based on an  astronomy catalog dataset obtained from the Sloan Digital Sky Survey (SDSS), a seismic dataset, as well as synthetic datasets. The SDSS catalog, published as part of the data release DR12, was downloaded from the project website link ($http://skyserver.sdss.org/CasJobs/$).\\ We consider a  projection of the dataset including attributes $(objID, ra, dec, u, g, r, i, z)$. The extracted dataset has a size of 800 MB containing around 6.7 million sky objects. The submitted query to obtain this dataset follows:

\begin{quote}
	\vspace{8pt}
	Select objID, ra, dec, u, g, r, i, z \newline
	From PhotoObjAll into MyTable
	\vspace{8pt}
\end{quote}

From the downloaded dataset, some subsets were extracted to produce datasets of different size. Additionally, in order to simulate very dense regions of the sky, we built
synthetic datasets with: 1000, 5000, 10000, 15000, and 20000 stars. The synthetic dataset includes millions of scaled solutions in a very dense region. Each solution is a multiplicative factor from a base query solution chosen uniformly within an interval of scale factors $s=[1.00000001,1.0000009]$. 

\subsubsection{Calibration}
We calibrated constellation query techniques using the SDSS dataset described above and a 3D seismic dataset from a region on the North Sea: Netherlands Offshore F3 Block – Complete \footnote{https://opendtect.org/osr/pmwiki.php/Main/Netherlands \newline OffshoreF3BlockComplete4GB}. The procedure aimed at finding the \emph{Einstein Cross} in the astronomy catalog and a seismic dome within the North Sea dataset, using our constellation query answering techniques. In both cases, the techniques succeeded in spotting the right structures among billions of candidates.

\subsubsection{Computing Environment}
 
The Constellation Query processing is implemented as an Apache Spark dataflow running on a shared nothing cluster.
The Petrus.lncc.br cluster is composed of 6 DELL PE R530 servers running CENTOS v. 7.2, kernel version $3.10.0\-327.13.1\\.el7.x86\_64$. Each cluster node includes a 2 Intel Xeon E5-2630 V3 2.4GHz processors, with 8 cores each, 96 GB of RAM memory, 20MB cache and 2 TB of hard disk. We are running Hadoop/HDFS v2.7.3, Spark v2.0.0 and Python v2.6. Spark was configured with 50 executors each running with 5GB of RAM memory and 1 core. The driver module was configured with 80GB of RAM memory.
The implementation builds the quadtree at the master node, at the driver module, and distributes the list of nodes at the tree entry level (see line 4 at Function ExecuteQuery in \ref{alg:QP}). Each worker node  then runs the \emph{property\_matching} and \emph{distance\_matching} algorithms. Finally, answers are collected in a single solution file.
 
\subsection{Quadtree Properties}
\begin{figure}[!ht]
	\vspace{8pt}
	\centering	\includegraphics[width=\linewidth]{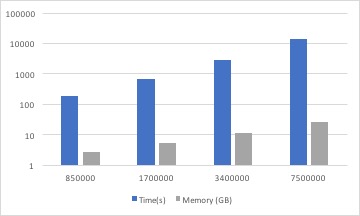}
	\caption{ Quadtree memory footprint and entry level access time}
	\label{fig:quadtreememory}
	\vspace{8pt}
\end{figure}

The quadtree has been designed to incur a low in-memory footprint. During construction, each node holds a stack with the stars covered by the respective geometric quadrant. Thus, when a split occurs, stars in the node stack are popped out and inserted into the stack of the children node covering its region. The quadtree construction continues until the entry tree level is achieved or the number of stars is less than 3. Only the leaves hold star data. Additionally, each node holds its geometric information, specified by its quadrant coordinates, including its geometric centroid. The initial tree height for large datasets is dependent on the \emph{additive factor} and not its data size. For large datasets, this means that the quadtree memory footprint should be proportional to the input dataset. As depicted in Figure \ref{fig:quadtreememory}, memory allocation(in GB) and the elapsed-time (in seconds) to retrieve all nodes at the tree entry level grow linearly with the dataset size. Observe that each raw entry size of the projected catalog is 44 bytes, which for an 850.000 stars would add up to approximately 37.4 MB. The actual quadtree memory allocation is 2.66 GB. 

\subsection{The Effectiveness of the Descent Tree algorithm}
The quadtree structure enables reducing the cost of constellation query processing by restricting composition computation to stars in pairs of nodes whose spatial quadrants match in distance. Selected matching pair nodes are evaluated for further splitting, according to cost model. In this section, we investigate the efficiency of the algorithm. We compare the cost of evaluating the stars matching  at the tree entry level with one that descends based on the cost model.

\begin{figure}[!ht]
	\vspace{8pt}
	\centering	\includegraphics[width=\linewidth]{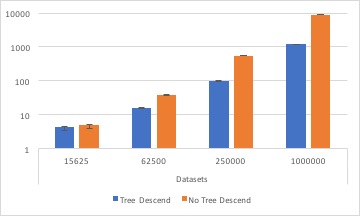}
	\caption{ Cost model efficiency in number  of comparisons}
	\label{fig:treedescendefficiency}
	\vspace{8pt}
\end{figure}

We ran the \emph{buildQuadtree} function with dense datasets and measured the difference in elapsed-time in both scenarios. Figure \ref{fig:treedescendefficiency} depicts the results on logarithm scale. The confidence interval is very small, between 0.04 to 36.86.
In terms of number of comparisons for 1 million stars, the cost model saves approximately 1.9x, leading to an order of magnitude on execution time savings.  

\subsection{Composition Algorithm Selection}


Once the matching stars have been paired and stored in the array of buckets, line 8  of Function \emph{ExecuteQuery} in algorithm \ref{alg:QP}, the compose algorithm \ref{alg:composition} applies a pairwise matching composition algorithm to compute solutions. In this section, we discuss the characteristics of the proposed composition algorithms.

In the first experiment a constellation query based on the Einstein cross elements is run for each composition algorithm and their elapsed-times are compared. The elapsed-time values correspond to the average of 10 runs measuring the maximum among all parallel execution nodes in each run. Additionally, we inform a confidence interval as $ conf \  =\  \alpha * \frac{\sigma}{\sqrt{n}}$ , where $\alpha\ is \ (1\ -\ confidence\  level)$, $\sigma $ is the standard deviation and $n$ is the number of runs in the experiment.


The geometric nature of constellation queries and the density of astronomical catalogs make the distance additive factor $\epsilon$ a very important element in query definition. As our experiments have shown, variations in this parameter may change a null result set to one with million of solutions.
So, it is unsurprising that the  additive factor is the main element influencing the proposed algorithms' computational costs. The experiments evaluate two classes of composition algorithms. In one class, we use the \emph{ Bucket\_NL} algorithm and, in the second one, we include the adoption of various \emph{Matrix Multiplication}  filtering strategies. 

\begin{figure}[!ht]
	\vspace{8pt}
	\centering
	\includegraphics[width=\linewidth]{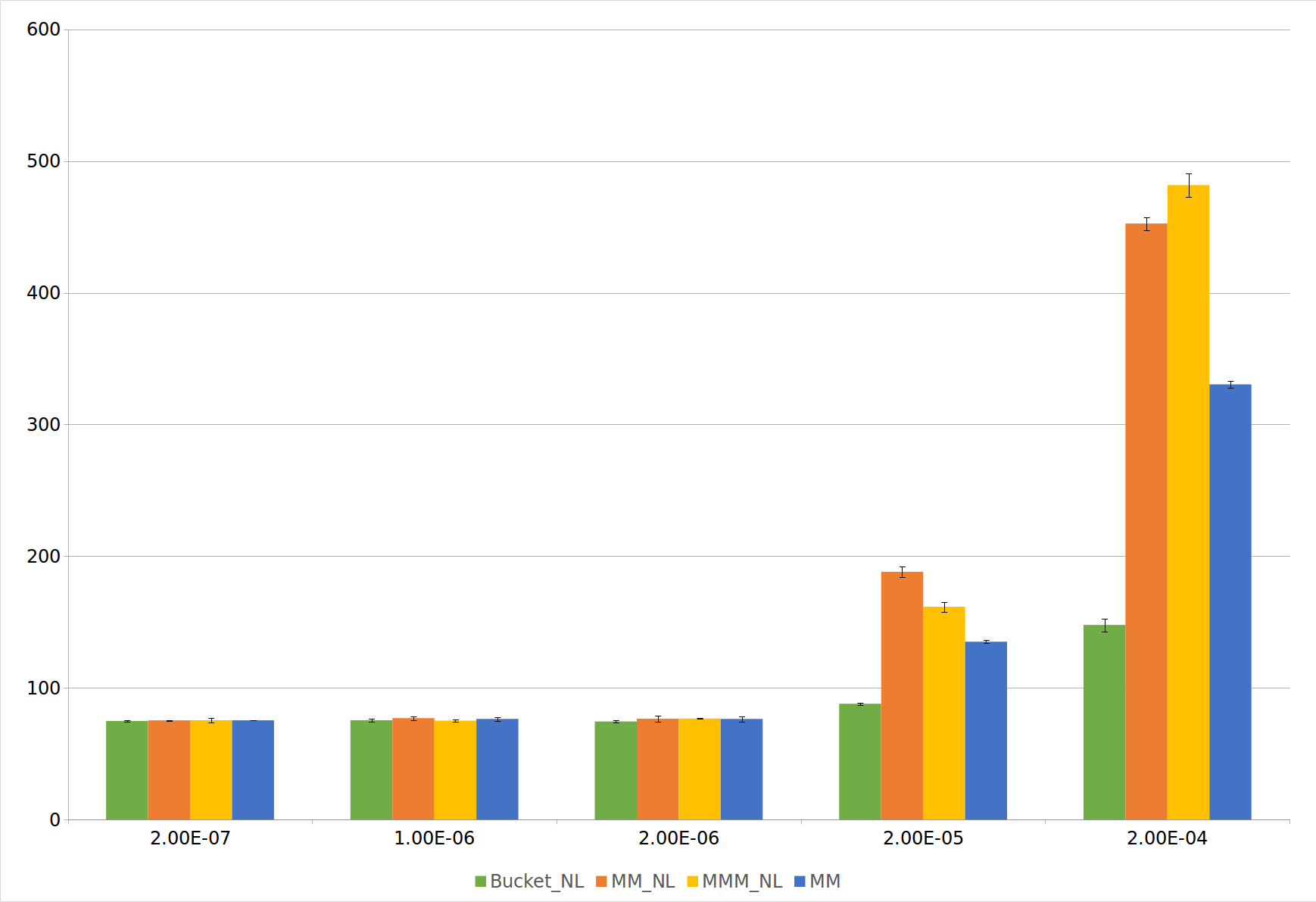}
	\caption{ Low Additive Error Tolerance $\epsilon$}
	\label{fig:low}
	\vspace{8pt}
\end{figure}

\begin{figure}[!ht]
	\vspace{8pt}
	\centering
	\includegraphics[width=\linewidth]{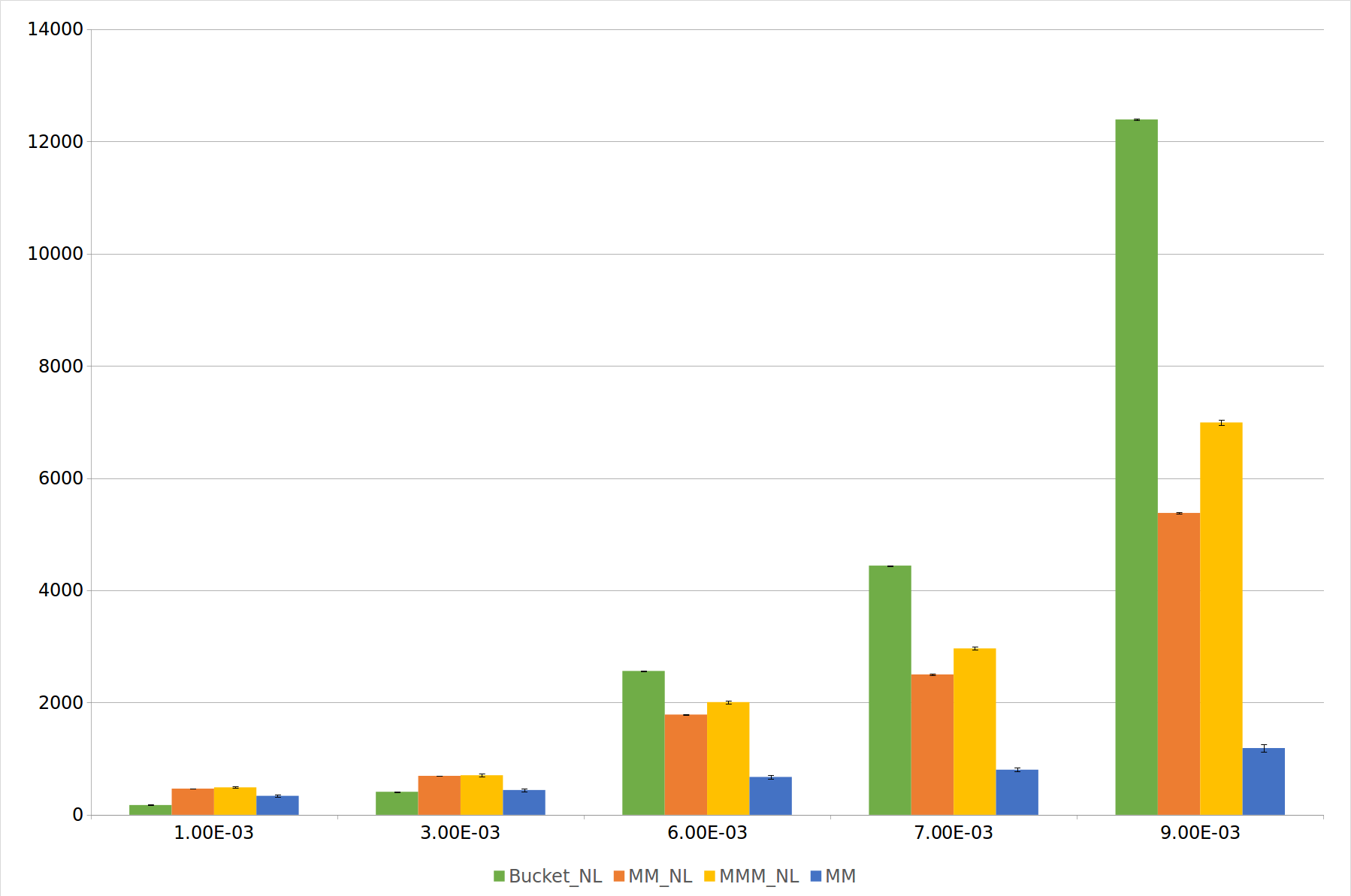}
	\caption{ High Additive Error Tolerance $\epsilon$. }
	\label{fig:high}
	\vspace{8pt}
\end{figure}

The experiment results are depicted in Figures \ref{fig:low} and \ref{fig:high}. In these plots, the horizontal axis presents different error tolerance values $\epsilon$,  while the vertical axis shows the elapsed-time of solving the constellation query using one of the composition algorithms.

Figure \ref{fig:low} shows basically two scenarios. For very small $\epsilon$, $ \le 10^{-6}$, the number of candidate elements in buckets is close to zero, leading to a total of 32 anchor elements to be selected and producing 52 candidate shapes.  In this scenario, the choice of a composition algorithms is irrelevant, with a difference in elapsed-time of less than $10\%$ among them. It is important to observe, however, that such a very restrictive constraint may eliminate interesting sets of stars. Unless the user is quite certain about the actual shape of its constellation, it is better to loosen the constraint.

The last blocks of runs involving the composition algorithms in Figure \ref{fig:low} shows that the results are different when  increasing $\epsilon$ by up to a factor of 100. Considering $\epsilon = 2,0 \times 10^{-5}$,we obtain $522,578$ productive anchor elements and an average of close to one element per bucket. The total number of candidate shapes rises to 12.6 million. In this setting, $Bucket\_NL$ is very fast, as it loops over very few elements in the  buckets to discover solutions.  The overhead of computing matrix multiplication is high, so  $Bucket\_NL$ is a clear winner. This scenario continues to hold up to $\epsilon = 2 \times 10^{-4}$, see Figure \ref{fig:low}. In this range, $Bucket\_NL$ is faster than $MM\_NL$ and $MMM\_NL$ by $214\%$ and  $240\%$, respectively.

Figure \ref{fig:high} highlights the behavior of algorithms under additive error tolerance values. The flexibility introduced by $\epsilon\  =\  $  $6 \times 10^{-3}$ generates $6.7$ million productive anchor elements and a total of $7.1$ billion solutions, with average elements per bucket of 10. In this scenario, eliminating non-productive anchor elements, close to $300,000$, by filtering using matrix multiplication eliminates the need of computing nested loops over approximately $405$ candidate elements in buckets. Thus, running fast matrix multiplication as a pre-step to nested-loop becomes beneficial. 
  
\begin{figure}[!ht]
	\vspace{8pt}
	\centering
	\includegraphics[width=\linewidth]{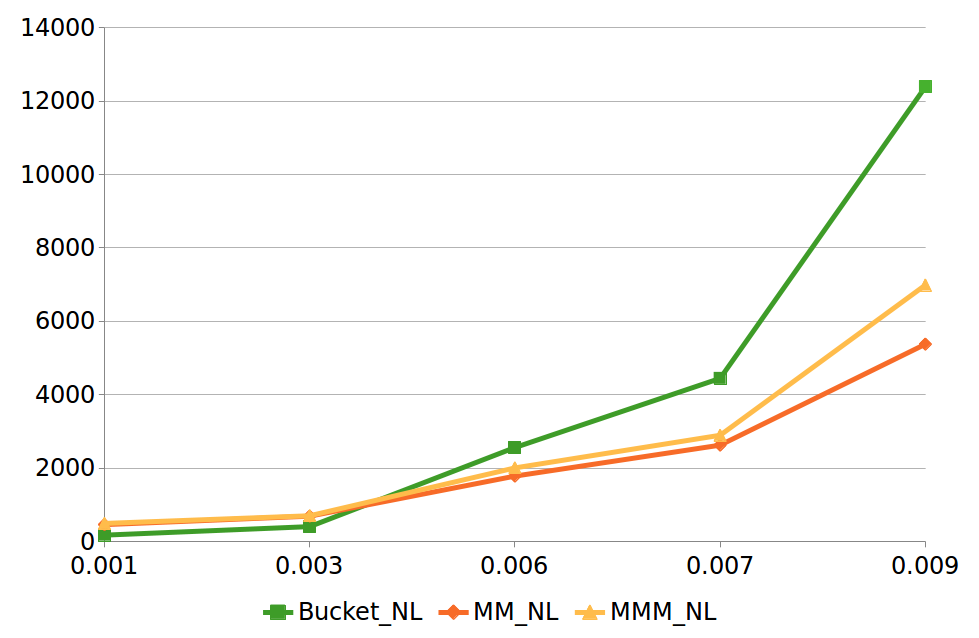}
	\caption{Zoom In on Matrix Multiplication: large threshold }
	\label{fig:MM}
	\vspace{8pt}
\end{figure}

Figure \ref{fig:MM} shows the  point at which matrix multiplication becomes beneficial: when $\epsilon = 6.0 \times 10^{-3}$  matrix multiplication starts to efficiently filter out anchor nodes and so the reduction in nested-loop time compensates for the cost of performing matrix multiplication. 
The gains observed by running matrix multiplication algorithms as a pre-filtering step before nested loop for $\epsilon$ in range $6 \times 10^{-3}$ and $9 \times 10^{-3}$ are up to $45.6\%$ for \emph{MM\_NL} and $34.6\%$ for \emph{MMM\_NL}.

%

\begin{figure}[!ht]
	\vspace{8pt}
	\centering
	\includegraphics[width=\linewidth]{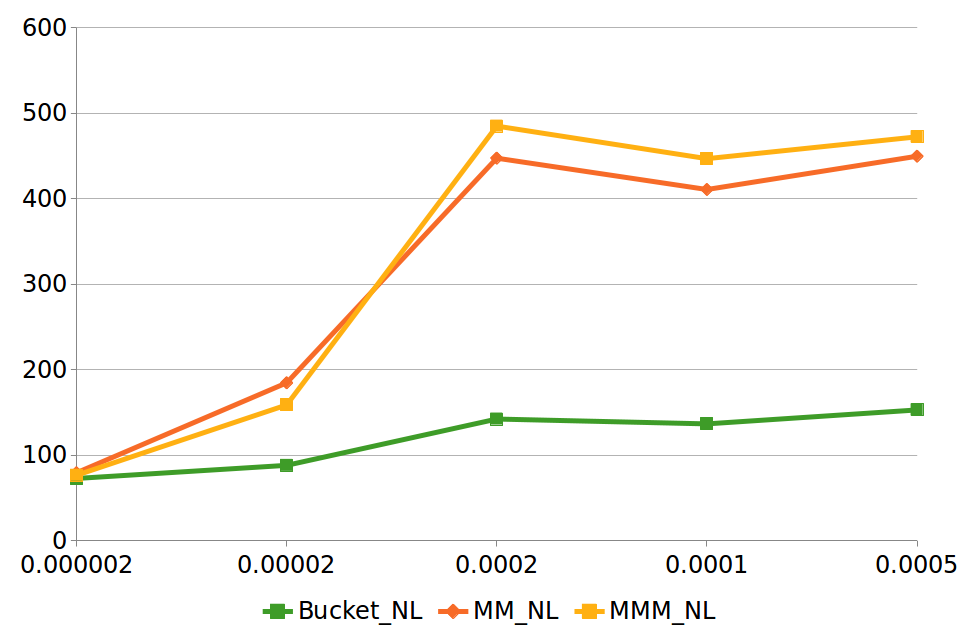}
	\caption{Zoom In on Matrix Multiplication: low threshold}
	\label{fig:MM_MMM}
	\vspace{8pt}
\end{figure}

 Figures \ref{fig:MM} and \ref{fig:MM_MMM} zoom in on the \emph{Matrix Multiplication} algorithms. The former shows the results with thresholds not less than $ 1 \times 10^{-3}$. In this range, we can observe an inversion in performance between $MM\_NL$ and $MMM\_NL$. The inflexion point occurs after $\epsilon \ge 2 \times 10^{-3}$. Threshold values below the inflexion point include anchor nodes with very few elements in buckets. In this scenario, computing multiple matrix multiplication is very fast. Moreover, elements that appear with zeros in the resulting matrix diagonal can be looked up in buckets and deleted, before the final nested-loop. The result is a gain of up to $14\%$ in elapsed-time with respect to $MM\_NL$. From the inflexion point on, $Matrix\_Multiplication\_NL$ is the best choice with gains up to $30\%$ with respect to $MMM\_NL$. The selection among composition algorithms is summarized in  Table \ref{tab:composition}, according to the results on the SDSS dataset.

\begin{table}[!ht]
	\vspace{8pt}
	\centering
	\caption{Composition Algorithms Selection}
	\begin{tabular}{ C {3cm} C{3cm}  }
		\hline\noalign{\smallskip}
		Threshold-Range & Best Choice \ Composition Algorithm  \\
		\hline\noalign{\smallskip}
		$ \le 0.003$ & $Bucket\_NL$  \\
		$ > 0.003$ &  $MM\_NL$   \\
		\hline\noalign{\smallskip}
	\end{tabular}
	\label{tab:composition}
	\vspace{8pt}
\end{table}

Finally, the matrix multiplication \emph{MM} algorithm is, as expected, a good choice for existential constellation queries which ask whether any subset of the dataset matches the query but does not ask to specify that subset.  In this scenario, once the matrix multiplication indicates a resulting matrix diagonal with all zeros, the anchor element produces no candidate shape and can be eliminated from the existential query result.

\subsection{Pure Constellation Scale-up}
\begin{figure}[!ht]
	\vspace{8pt}
	\centering
	\includegraphics[width=\linewidth]{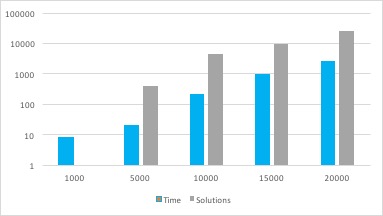}
	\caption{Pure Constellation Scale-up}
	\label{fig:PC_SUP}
	\vspace{8pt}
\end{figure}

We investigated Pure CQ scale-up adopting the set of dense datasets (see section \ref{subsec:dataset_configuration}), error bound epsilon = $4.4 x 10^{-6}$ and Bucket\_NL for the composition algorithm. The execution produced solutions of size: zero, 21, 221, 1015, and 2685. The run with 1000 stars dataset produced zero solutions, which shows the relevance of tunning the error bound for a given dataset and the restrictions imposed by Pure CQ. Apart from the runs with the 15,000 stars dataset, the variations in time followed the increase in the number of solutions.This indicates that non solutions are quickly discarded and the time is mostly due to producing solutions. Figure \ref{fig:PC_SUP} depicts the results, where time corresponds to the elapsed-time in seconds of the parallel execution.

\subsection{Assessing General Constellation Queries}
In this last experiment, we assess General CQ. We have run the algorithm using a dense dataset with 1000 stars and have configured the system to accept scale factors between 0 and 5 and to consider an error bound of $\epsilon=4.4x10^{-6}$. General CQ produces a huge number of solutions within the provided scale interval. Our experiments produced 160 million results and executed for 25 minutes, for an average of 10 runs. 

The results show that the pruning techniques efficiently reduced the total number of solutions that could have reached in the billions. The number of solutions and the time depend on both the error bound and the scale interval. As a test of this effect, we ran General CQ with 5000 stars dataset allowing the scales to range between 0.5 and 1. The execution  took 15:45 minutes and produced 50.17 million solutions. 

\section{Related Work}
\label{sec:related_works}

Finding collections of objects having some metric relationship of interest  is an area with many applications. The problem has different names depending on the discipline, including \emph{Object Identification} \cite{ObjectIdentification}, \emph{Graph Queries} \cite{zou_answering_2011},  \emph{Pattern Matching} and \emph{Pattern Recognition} \cite{PatternRecognition-MachineLearning}. 

Pattern recognition research focuses on  identifying patterns and regularities in data \cite{PatternRecognition-MachineLearning}. Graphs are commonly used in pattern recognition due to their flexibility in representing structural geometric and relational descriptions for concepts, such as pixels, predicates, and objects \cite{GraphMatchingHistory, GRAPHMATCHING-PATTERNRECOGNITION}. In this way, problems are commonly posed as a graph query problem, such as subgraph search, shortest-path query, reachability verification, and pattern match. Among these, subgraph matching queries are  related to our work. 

In a subgraph query, a query is a connected set of nodes and edges (which may or may not be labeled). A match is a (usually non-induced) subgraph of a large graph that is isomorphic to the query. While the literature in that field is vast [\cite{zou_distance-join:_2009},\cite{zou_answering_2011}, \cite{GiugnoS02}], the problem is fundamentally different, because there is no notion of space (so data structures like quadtrees are useless) and there is no distance notion of scale (the $\epsilon$ that plays such a big role for us).

Finally, constellation queries are a class of package queries (PQ), Brucato et al. \cite{Package2016}. A PQ query enables the definition of packages through local constraints on tuples, global constraints on packages of tuples, and an optimization function to rank packages in results. Using this formulation, a package query is translated into a linear programing problem and solved by a linear solver. Constellation Queries also produces packages but local and global constraints require labeling tuples, which in PQ leads to self joins in the number of elements of the query that would be impractical for large datasets. On the other hand, in order
to avoid impractical self-joins, we implement a full query processing strategy tailored to CQ that:(i) prune neighbors;(ii) reduces the number of comparisons; (iii) optimize global constraint using matrix multiplication.  The PQ \emph{Sketch Refine} algorithm relies on data partitioning and quadtree on representative tuples to reduce the initial number of packages. In CQ, we apply a cost model to guide quadtree descent to nodes with matching stars, reducing the number of comparisons by half. CQ implementation has been designed to run in parallel enabling processing of large astronomic datasets. Finally, general constellation queries offer a sound and complete strategy to retrieve solutions at any scale, which extends the power of CQ in finding unknown constellations. 

Table \ref{tab:comparison} summarizes the major characteristics of the above works as compared with Constellation Queries CQ. CQ and Package Queries share the same data complexity exposing a combinatorial behavior, requiring pruning techniques. Looking for patterns in graphs is a more constrained problem, in which relationships between nodes are preset by graph edges. In a scenario of a complete graph, (c) becomes approximately, $O(|D|^3 )$. One may easily observe that the complexity of CQ is orders of magnitude larger than graph pattern, which confirms the intuition that it is a less constrained composition problem. In this context, finding strategies that reduce the search space are even more vital than in the graph case.

\begin{table}[!ht]
	\vspace{8pt}
	\centering
	\caption{Comparison of Approaches}
	\begin{tabular}{ L{1.5cm} C{1cm} C{1.7cm} C{1.7cm}  }
		\hline\noalign{\smallskip}
		Approach \ Criteria & Basic \ Model & Object Composition & Complexity \\
		\hline
		CQ & lists &Shape matching & $O(\binom{|D|} {k})$ \\
		\\
        PQ & sets &Global constraints & $O(\binom{|D|} {k})$\\
		Graph Pattern & graph & Matching Constrained & 	$O(|D|*|E| + |E_{q}|*|D|^{2} + |Q|*|D|)$ \footnote{only within the \newline graph structure} \\
		\\
		\hline\noalign{\smallskip}
	\end{tabular}
	\label{tab:comparison}
	\vspace{8pt}
\end{table}

\section{Conclusion}
\label{sec:conclusion}

In this paper, we introduce \emph{constellation queries}, specified as a geometrical composition of individual elements from a big dataset. We illustrate the application of Constellation Queries in astronomy (e.g. Einstein crosses) and seismic data (the location of salt domes). 

The objective of a constellation query is to find lists of dataset elements that form a spatial structure geometrically similar to the one given by the query and whose corresponding elements share common attribute values with the ones from the query.
Answering a Constellation Query is hard due to the large number of candidate subsets possible $O(|D|^{|Q|})$ candidates, where $|D|$ is the size of the dataset and $|Q|$ is the size of the Constellation query. 

We have designed  procedures to efficiently compute both pure and general Constellation Queries. For pure constellation queries, first, we reduce the space of possible candidate sets by associating to each element in the dataset neighbors at a maximum distance, corresponding to the largest distance between any two elements in the query. Next, we filtered candidates yet further into buckets through the use of a quadtree. Next, we used a bucket joining algorithm, optionally preceded by a matrix multiplication filter to find solutions.

For general constellation queries, we are looking for sequences in the data set that match a pattern query based on a multiplicative scale factor. The main challenge is that general constellation queries must discover appropriate scale factors which can take any real value. Our discrete algorithm provably finds all sequences in the data set at every scale
that match  a pattern query within an error tolerance,
even though there are an uncountable infinite number of scales.

Our experiments run on Spark, running on the neighboring dataset distributed over HDFS. Our work shows that our filtering techniques having to do with quadtrees are enormously beneficial, whereas matrix multiplication is beneficial only in high density settings. 

To the best of our knowledge, this is the first work to investigate constellation queries. There are numerous opportunities for future work, especially in optimization for higher dimensions. 

\section{Acknowledgment}

This research is partially funded by EU H2020 Program and MCTI/RNP-Brazil(HPC4e Project - grant agreement number 689772),
FAPERJ (MUSIC Project E36-2013) and INRIA (SciDISC 2017), INRIA international chair, U.S. National Science Foundation MCB-1158273, IOS-1139362 and MCB-1412232. This support is greatly appreciated.

\bibliographystyle{ieeetr}
\bibliography{VLDB_Constellation}

\end{document}